\documentclass[12pt,a4paper]{article}
\usepackage{epsfig}
\font\ninerm=cmr9
\def\invpb{\ensuremath{{{\rm pb}^{-1}}}}

\def\Gc{\ensuremath{{\mathrm{GeV}/c}}}
\def\Gcs{\ensuremath{{\mathrm{GeV}/c^2}}}

\def\h{\ensuremath{{\mathrm h}}}
\def\A{\ensuremath{{\mathrm A}}}

\def\Z{\ensuremath{{\mathrm Z}}}
\def\W{\ensuremath{{\mathrm W}}}

\def\l{\ensuremath{{\ell}}}

\def\bq{\ensuremath{{\mathrm b}}}

\def\b{\bq}

\def\WW{\ensuremath{{\mathrm W}^{+}{\mathrm W}^{-}}}
\def\ZZ{\ensuremath{{\mathrm Z}{\mathrm Z}}}
\def\ee{\ensuremath{{\mathrm e}^{+}{\mathrm e}^{-}}}

\def\bb{\ensuremath{{\mathrm b}\bar{\mathrm b}}}
\def\cc{\ensuremath{{\mathrm c}\bar{\mathrm c}}}

\def\qq{\ensuremath{{\mathrm q}\bar{\mathrm q}}}
\def\qqbar{\qq}

\def\nunu{\mbox{\ensuremath{\nu\bar{\nu}}}}

\def\nnbar{\nunu}
\def\tptm{\mbox{\ensuremath{{\tau}^{+}{\tau}^{-}}}}

\def\lplm{\mbox{\ensuremath{{\l}^{+}{\l}^{-}}}}

\def\mh{\mbox{\ensuremath{m_{\h}}}}

\def\mA{\mbox{\ensuremath{m_{\A}}}}
\def\mZ{\mbox{\ensuremath{m_{\Z}}}}
\def\mtop{\mbox{\ensuremath{m_{\mathrm{top}}}}}
\def\mhmax{\mbox{\ensuremath{m_h^{\mathrm{max}}}}}
\def\to{\mbox{\ensuremath{ \rightarrow \ }}}
\def\r{\to}

\def\roots{\ensuremath{{\sqrt{s}}}}
\def\tanb{\ensuremath{\tan\beta}}
\def\sba{\ensuremath{\sin^2(\beta-\alpha)}}

\def\cba{\ensuremath{\cos^2(\beta-\alpha)}}

\def\mrec{\ensuremath{M_{\rm rec}}}

\newcommand{\ALEPH}{{ALEPH}}

\newcommand{\LEP}{{LEP}}

\newcommand{\CERN}{{CERN}}
\def\PLB#1#2#3{{ Phys. Lett. }{\bf B#1 }(#2) #3}

\def\EPJ#1#2#3{{ Eur. Phys. J. }{\bf C#1 }(#2) #3}
\def\NIM#1#2#3{{ Nucl. Instrum. and Methods}
{\bf A#1 }(#2) #3}

\begin{document}\setlength{\unitlength}{1mm}
%
%
%
\begin{titlepage}

\setlength{\topmargin}{0.5cm}
\setlength{\oddsidemargin}{-0.2cm}
\vspace{1cm}
\date{}
\title{
\mbox{ }\\
\vspace{1.0cm}
\mbox{ }\\
Final results of the searches for neutral Higgs bosons \\
in $\ee$ collisions at $\roots$ up to 209 GeV
}
\vskip 1cm
\author{The {\sc ALEPH} Collaboration $^{*)}$}
\vspace{1cm}
\maketitle
%
\begin{picture}(160,1)
\put(-5,95){\large EUROPEAN ORGANIZATION FOR NUCLEAR RESEARCH (CERN)}
\put(27,88){\rm \large European Laboratory for Particle Physics}
\put(120,70){\parbox[t]{45mm}{CERN-EP/2001-095}}
\put(120,64){\parbox[t]{45mm}{18-Dec-2001}}
\end{picture}
\thispagestyle{empty}
\begin{abstract}
The final results of the ALEPH search for the Standard Model Higgs
boson at LEP, with data collected in the year 2000 at centre-of-mass
energies up to 209\,GeV, are presented. The changes with respect to
the preceding publication are described and a complete study of
systematic effects is reported. The findings of this final analysis
confirm the preliminary results published in November 2000 shortly
after the closing down of the LEP collider: a significant excess of
events is observed, consistent with the production of a $115\,\Gcs$
Standard Model Higgs boson.
The final results of the searches for the neutral Higgs bosons of the
MSSM are also reported, in terms of limits on $\mh$, $\mA$ and
$\tanb$.
Limits are also set on $\mh$ in the case of invisible decays.

\end{abstract}

\vfill
\centerline{\it To be published in Physics Letters B}
\vskip .5cm
\noindent
--------------------------------------------\hfil\break
{\ninerm $^*$) See next pages for the list of authors.}
\end{titlepage}

\pagestyle{empty}
\newpage
\small
%
%
\newlength{\saveparskip}
\newlength{\savetextheight}
\newlength{\savetopmargin}
\newlength{\savetextwidth}
\newlength{\saveoddsidemargin}
\newlength{\savetopsep}
\setlength{\saveparskip}{\parskip}
\setlength{\savetextheight}{\textheight}
\setlength{\savetopmargin}{\topmargin}
\setlength{\savetextwidth}{\textwidth}
\setlength{\saveoddsidemargin}{\oddsidemargin}
\setlength{\savetopsep}{\topsep}
%
%
\setlength{\parskip}{0.0cm}
\setlength{\textheight}{25.0cm} 
\setlength{\topmargin}{-1.5cm}
\setlength{\textwidth}{16 cm}
\setlength{\oddsidemargin}{-0.0cm}
\setlength{\topsep}{1mm}
\pretolerance=10000
\centerline{\large\bf The ALEPH Collaboration}
\footnotesize
\vspace{0.5cm}
{\raggedbottom
\begin{sloppypar}
\samepage\noindent
A.~Heister,
S.~Schael
\nopagebreak
\begin{center}
\parbox{15.5cm}{\sl\samepage
Physikalisches Institut das RWTH-Aachen, D-52056 Aachen, Germany}
\end{center}\end{sloppypar}
\vspace{2mm}
\begin{sloppypar}
\noindent
R.~Barate,
R.~Bruneli\`ere,
I.~De~Bonis,
D.~Decamp,
C.~Goy,
S.~Jezequel,
J.-P.~Lees,
F.~Martin,
E.~Merle,
\mbox{M.-N.~Minard},
B.~Pietrzyk,
B.~Trocm\'e
\nopagebreak
\begin{center}
\parbox{15.5cm}{\sl\samepage
Laboratoire de Physique des Particules (LAPP), IN$^{2}$P$^{3}$-CNRS,
F-74019 Annecy-le-Vieux Cedex, France}
\end{center}\end{sloppypar}
\vspace{2mm}
\begin{sloppypar}
\noindent
G.~Boix,
S.~Bravo,
M.P.~Casado,
M.~Chmeissani,
J.M.~Crespo,
E.~Fernandez,
M.~Fernandez-Bosman,
Ll.~Garrido,$^{15}$
E.~Graug\'{e}s,
J.~Lopez,
M.~Martinez,
G.~Merino,
R.~Miquel,$^{31}$
Ll.M.~Mir,$^{31}$
A.~Pacheco,
D.~Paneque,
H.~Ruiz
\nopagebreak
\begin{center}
\parbox{15.5cm}{\sl\samepage
Institut de F\'{i}sica d'Altes Energies, Universitat Aut\`{o}noma
de Barcelona, E-08193 Bellaterra (Barcelona), Spain$^{7}$}
\end{center}\end{sloppypar}
\vspace{2mm}
\begin{sloppypar}
\noindent
A.~Colaleo,
D.~Creanza,
N.~De~Filippis,
M.~de~Palma,
G.~Iaselli,
G.~Maggi,
M.~Maggi,
S.~Nuzzo,
A.~Ranieri,
G.~Raso,$^{24}$
F.~Ruggieri,
G.~Selvaggi,
L.~Silvestris,
P.~Tempesta,
A.~Tricomi,$^{3}$
G.~Zito
\nopagebreak
\begin{center}
\parbox{15.5cm}{\sl\samepage
Dipartimento di Fisica, INFN Sezione di Bari, I-70126 Bari, Italy}
\end{center}\end{sloppypar}
\vspace{2mm}
\begin{sloppypar}
\noindent
X.~Huang,
J.~Lin,
Q. Ouyang,
T.~Wang,
Y.~Xie,
R.~Xu,
S.~Xue,
J.~Zhang,
L.~Zhang,
W.~Zhao
\nopagebreak
\begin{center}
\parbox{15.5cm}{\sl\samepage
Institute of High Energy Physics, Academia Sinica, Beijing, The People's
Republic of China$^{8}$}
\end{center}\end{sloppypar}
\vspace{2mm}
\begin{sloppypar}
\noindent
D.~Abbaneo,
P.~Azzurri,
T.~Barklow,$^{30}$
O.~Buchm\"uller,$^{30}$
M.~Cattaneo,
F.~Cerutti,
B.~Clerbaux,
H.~Drevermann,
R.W.~Forty,
M.~Frank,
F.~Gianotti,
T.C.~Greening,$^{26}$
J.B.~Hansen,
J.~Harvey,
D.E.~Hutchcroft,
P.~Janot,
B.~Jost,
M.~Kado,$^{31}$
P.~Maley,
P.~Mato,
A.~Moutoussi,
F.~Ranjard,
L.~Rolandi,
D.~Schlatter,
G.~Sguazzoni,
W.~Tejessy,
F.~Teubert,
A.~Valassi,
I.~Videau,
J.J.~Ward
\nopagebreak
\begin{center}
\parbox{15.5cm}{\sl\samepage
European Laboratory for Particle Physics (CERN), CH-1211 Geneva 23,
Switzerland}
\end{center}\end{sloppypar}
\vspace{2mm}
\begin{sloppypar}
\noindent
F.~Badaud,
S.~Dessagne,
A.~Falvard,$^{20}$
D.~Fayolle,
P.~Gay,
J.~Jousset,
B.~Michel,
S.~Monteil,
D.~Pallin,
J.M.~Pascolo,
P.~Perret
\nopagebreak
\begin{center}
\parbox{15.5cm}{\sl\samepage
Laboratoire de Physique Corpusculaire, Universit\'e Blaise Pascal,
IN$^{2}$P$^{3}$-CNRS, Clermont-Ferrand, F-63177 Aubi\`{e}re, France}
\end{center}\end{sloppypar}
\vspace{2mm}
\begin{sloppypar}
\noindent
J.D.~Hansen,
J.R.~Hansen,
P.H.~Hansen,
B.S.~Nilsson,
A.~W\"a\"an\"anen
\nopagebreak
\begin{center}
\parbox{15.5cm}{\sl\samepage
Niels Bohr Institute, 2100 Copenhagen, DK-Denmark$^{9}$}
\end{center}\end{sloppypar}
\vspace{2mm}
\begin{sloppypar}
\noindent
A.~Kyriakis,
C.~Markou,
E.~Simopoulou,
A.~Vayaki,
K.~Zachariadou
\nopagebreak
\begin{center}
\parbox{15.5cm}{\sl\samepage
Nuclear Research Center Demokritos (NRCD), GR-15310 Attiki, Greece}
\end{center}\end{sloppypar}
\vspace{2mm}
\begin{sloppypar}
\noindent
A.~Blondel,$^{12}$
\mbox{J.-C.~Brient},
F.~Machefert,
A.~Roug\'{e},
M.~Swynghedauw,
R.~Tanaka
\linebreak
H.~Videau
\nopagebreak
\begin{center}
\parbox{15.5cm}{\sl\samepage
Laboratoire de Physique Nucl\'eaire et des Hautes Energies, Ecole
Polytechnique, IN$^{2}$P$^{3}$-CNRS, \mbox{F-91128} Palaiseau Cedex, France}
\end{center}\end{sloppypar}
\vspace{2mm}
\begin{sloppypar}
\noindent
V.~Ciulli,
E.~Focardi,
G.~Parrini
\nopagebreak
\begin{center}
\parbox{15.5cm}{\sl\samepage
Dipartimento di Fisica, Universit\`a di Firenze, INFN Sezione di Firenze,
I-50125 Firenze, Italy}
\end{center}\end{sloppypar}
\vspace{2mm}
\begin{sloppypar}
\noindent
A.~Antonelli,
M.~Antonelli,
G.~Bencivenni,
G.~Bologna,$^{4}$
F.~Bossi,
P.~Campana,
G.~Capon,
V.~Chiarella,
P.~Laurelli,
G.~Mannocchi,$^{5}$
F.~Murtas,
G.P.~Murtas,
L.~Passalacqua,
M.~Pepe-Altarelli,$^{25}$
P.~Spagnolo
\nopagebreak
\begin{center}
\parbox{15.5cm}{\sl\samepage
Laboratori Nazionali dell'INFN (LNF-INFN), I-00044 Frascati, Italy}
\end{center}\end{sloppypar}
\vspace{2mm}
\begin{sloppypar}
\noindent
J.~Kennedy,
J.G.~Lynch,
P.~Negus,
V.~O'Shea,
D.~Smith,
A.S.~Thompson
\nopagebreak
\begin{center}
\parbox{15.5cm}{\sl\samepage
Department of Physics and Astronomy, University of Glasgow, Glasgow G12
8QQ,United Kingdom$^{10}$}
\end{center}\end{sloppypar}
\vspace{2mm}
\pagebreak 
\begin{sloppypar}
\noindent
S.~Wasserbaech
\nopagebreak
\begin{center}
\parbox{15.5cm}{\sl\samepage
Department of Physics, Haverford College, Haverford, PA 19041-1392, U.S.A.}
\end{center}\end{sloppypar}
\vspace{2mm}
\begin{sloppypar}
\noindent
R.~Cavanaugh,
S.~Dhamotharan,
C.~Geweniger,
P.~Hanke,
V.~Hepp,
E.E.~Kluge,
G.~Leibenguth,
A.~Putzer,
H.~Stenzel,
K.~Tittel,
S.~Werner,$^{19}$
M.~Wunsch$^{19}$
\nopagebreak
\begin{center}
\parbox{15.5cm}{\sl\samepage
Kirchhoff-Institut f\"ur Physik, Universit\"at Heidelberg, D-69120
Heidelberg, Germany$^{16}$}
\end{center}\end{sloppypar}
\vspace{2mm}
\begin{sloppypar}
\noindent
R.~Beuselinck,
D.M.~Binnie,
W.~Cameron,
G.~Davies,
P.J.~Dornan,
M.~Girone,$^{1}$
R.D.~Hill,
N.~Marinelli,
J.~Nowell,
H.~Przysiezniak,$^{2}$
S.A.~Rutherford,
J.K.~Sedgbeer,
J.C.~Thompson,$^{14}$
R.~White
\nopagebreak
\begin{center}
\parbox{15.5cm}{\sl\samepage
Department of Physics, Imperial College, London SW7 2BZ,
United Kingdom$^{10}$}
\end{center}\end{sloppypar}
\vspace{2mm}
\begin{sloppypar}
\noindent
V.M.~Ghete,
P.~Girtler,
E.~Kneringer,
D.~Kuhn,
G.~Rudolph
\nopagebreak
\begin{center}
\parbox{15.5cm}{\sl\samepage
Institut f\"ur Experimentalphysik, Universit\"at Innsbruck, A-6020
Innsbruck, Austria$^{18}$}
\end{center}\end{sloppypar}
\vspace{2mm}
\begin{sloppypar}
\noindent
E.~Bouhova-Thacker,
C.K.~Bowdery,
D.P.~Clarke,
G.~Ellis,
A.J.~Finch,
F.~Foster,
G.~Hughes,
R.W.L.~Jones,
M.R.~Pearson,
N.A.~Robertson,
M.~Smizanska
\nopagebreak
\begin{center}
\parbox{15.5cm}{\sl\samepage
Department of Physics, University of Lancaster, Lancaster LA1 4YB,
United Kingdom$^{10}$}
\end{center}\end{sloppypar}
\vspace{2mm}
\begin{sloppypar}
\noindent
V.~Lemaitre
\nopagebreak
\begin{center}
\parbox{15.5cm}{\sl\samepage
Institut de Physique Nucl\'eaire, D\'epartement de Physique, Universit\'e Catholique de Louvain, 1348 Louvain-la-Neuve, Belgium}
\end{center}\end{sloppypar}
\vspace{2mm}
\begin{sloppypar}
\noindent
U.~Blumenschein,
F.~H\"olldorfer,
K.~Jakobs,
F.~Kayser,
K.~Kleinknecgt,
A.-S.~M\"uller,
G.~Quast,$^{6}$
B.~Renk,
H.-G.~Sander,
S.~Schmeling,
H.~Wachsmuth,
C.~Zeitnitz,
T.~Ziegler
\nopagebreak
\begin{center}
\parbox{15.5cm}{\sl\samepage
Institut f\"ur Physik, Universit\"at Mainz, D-55099 Mainz, Germany$^{16}$}
\end{center}\end{sloppypar}
\vspace{2mm}
\begin{sloppypar}
\noindent
A.~Bonissent,
J.~Carr,
P.~Coyle,
C.~Curtil,
A.~Ealet,
D.~Fouchez,
O.~Leroy,
T.~Kachelhoffer,
P.~Payre,
D.~Rousseau,
A.~Tilquin
\nopagebreak
\begin{center}
\parbox{15.5cm}{\sl\samepage
Centre de Physique des Particules de Marseille, Univ M\'editerran\'ee,
IN$^{2}$P$^{3}$-CNRS, F-13288 Marseille, France}
\end{center}\end{sloppypar}
\vspace{2mm}
\begin{sloppypar}
\noindent
F.~Ragusa
\nopagebreak
\begin{center}
\parbox{15.5cm}{\sl\samepage
Dipartimento di Fisica, Universit\`a di Milano e INFN Sezione di
Milano, I-20133 Milano, Italy.}
\end{center}\end{sloppypar}
\vspace{2mm}
\begin{sloppypar}
\noindent
A.~David,
H.~Dietl,
G.~Ganis,$^{27}$
K.~H\"uttmann,
G.~L\"utjens,
C.~Mannert,
W.~M\"anner,
\mbox{H.-G.~Moser},
R.~Settles,
G.~Wolf
\nopagebreak
\begin{center}
\parbox{15.5cm}{\sl\samepage
Max-Planck-Institut f\"ur Physik, Werner-Heisenberg-Institut,
D-80805 M\"unchen, Germany\footnotemark[16]}
\end{center}\end{sloppypar}
\vspace{2mm}
\begin{sloppypar}
\noindent
J.~Boucrot,
O.~Callot,
M.~Davier,
L.~Duflot,
\mbox{J.-F.~Grivaz},
Ph.~Heusse,
A.~Jacholkowska,$^{20}$
C.~Loomis,
L.~Serin,
\mbox{J.-J.~Veillet},
J.-B.~de~Vivie~de~R\'egie,$^{28}$
C.~Yuan
\nopagebreak
\begin{center}
\parbox{15.5cm}{\sl\samepage
Laboratoire de l'Acc\'el\'erateur Lin\'eaire, Universit\'e de Paris-Sud,
IN$^{2}$P$^{3}$-CNRS, F-91898 Orsay Cedex, France}
\end{center}\end{sloppypar}
\vspace{2mm}
\begin{sloppypar}
\noindent
G.~Bagliesi,
T.~Boccali,
L.~Fo\`a,
A.~Giammanco,
A.~Giassi,
F.~Ligabue,
A.~Messineo,
F.~Palla,
G.~Sanguinetti,
A.~Sciab\`a,
R.~Tenchini,$^{1}$
A.~Venturi,$^{1}$
P.G.~Verdini
\samepage
\begin{center}
\parbox{15.5cm}{\sl\samepage
Dipartimento di Fisica dell'Universit\`a, INFN Sezione di Pisa,
e Scuola Normale Superiore, I-56010 Pisa, Italy}
\end{center}\end{sloppypar}
\vspace{2mm}
\begin{sloppypar}
\noindent
O.~Awunor,
G.A.~Blair,
J.~Coles,
G.~Cowan,
A.~Garcia-Bellido,
M.G.~Green,
L.T.~Jones,
T.~Medcalf,
A.~Misiejuk,
J.A.~Strong,
P.~Teixeira-Dias
\nopagebreak
\begin{center}
\parbox{15.5cm}{\sl\samepage
Department of Physics, Royal Holloway \& Bedford New College,
University of London, Egham, Surrey TW20 OEX, United Kingdom$^{10}$}
\end{center}\end{sloppypar}
\vspace{2mm}
\begin{sloppypar}
\noindent
R.W.~Clifft,
T.R.~Edgecock,
P.R.~Norton,
I.R.~Tomalin
\nopagebreak
\begin{center}
\parbox{15.5cm}{\sl\samepage
Particle Physics Dept., Rutherford Appleton Laboratory,
Chilton, Didcot, Oxon OX11 OQX, United Kingdom$^{10}$}
\end{center}\end{sloppypar}
\vspace{2mm}
\begin{sloppypar}
\noindent
\mbox{B.~Bloch-Devaux},
D.~Boumediene,
P.~Colas,
B.~Fabbro,
E.~Lan\c{c}on,
\mbox{M.-C.~Lemaire},
E.~Locci,
P.~Perez,
J.~Rander,
\mbox{J.-F.~Renardy},
A.~Rosowsky,
P.~Seager,$^{13}$
A.~Trabelsi,$^{21}$
B.~Tuchming,
B.~Vallage
\nopagebreak
\begin{center}
\parbox{15.5cm}{\sl\samepage
CEA, DAPNIA/Service de Physique des Particules,
CE-Saclay, F-91191 Gif-sur-Yvette Cedex, France$^{17}$}
\end{center}\end{sloppypar}
\setlength{\topsep}{-15mm} 
\vspace{2mm}
\setlength{\textheight}{24.0cm} 
\pagebreak 
\setlength{\topsep}{1mm} 
\begin{sloppypar}
\noindent
N.~Konstantinidis,
A.M.~Litke,
G.~Taylor
\nopagebreak
\begin{center}
\parbox{15.5cm}{\sl\samepage
Institute for Particle Physics, University of California at
Santa Cruz, Santa Cruz, CA 95064, USA$^{22}$}
\end{center}\end{sloppypar}
\vspace{2mm}
\begin{sloppypar}
\noindent
C.N.~Booth,
S.~Cartwright,
F.~Combley,$^{4}$
P.N.~Hodgson,
M.~Lehto,
L.F.~Thompson
\nopagebreak
\begin{center}
\parbox{15.5cm}{\sl\samepage
Department of Physics, University of Sheffield, Sheffield S3 7RH,
United Kingdom$^{10}$}
\end{center}\end{sloppypar}
\vspace{2mm}
\begin{sloppypar}
\noindent
K.~Affholderbach,$^{23}$
A.~B\"ohrer,
S.~Brandt,
C.~Grupen,
J.~Hess,
A.~Ngac,
G.~Prange,
U.~Sieler
\nopagebreak
\begin{center}
\parbox{15.5cm}{\sl\samepage
Fachbereich Physik, Universit\"at Siegen, D-57068 Siegen, Germany$^{16}$}
\end{center}\end{sloppypar}
\vspace{2mm}
\begin{sloppypar}
\noindent
C.~Borean,
G.~Giannini
\nopagebreak
\begin{center}
\parbox{15.5cm}{\sl\samepage
Dipartimento di Fisica, Universit\`a di Trieste e INFN Sezione di Trieste,
I-34127 Trieste, Italy}
\end{center}\end{sloppypar}
\vspace{2mm}
\begin{sloppypar}
\noindent
H.~He,
J.~Putz,
J.~Rothberg
\nopagebreak
\begin{center}
\parbox{15.5cm}{\sl\samepage
Experimental Elementary Particle Physics, University of Washington, Seattle,
WA 98195 U.S.A.}
\end{center}\end{sloppypar}
\vspace{2mm}
\begin{sloppypar}
\noindent
S.R.~Armstrong,
K.~Berkelman,
K.~Cranmer,
D.P.S.~Ferguson,
Y.~Gao,$^{29}$
S.~Gonz\'{a}lez,
O.J.~Hayes,
H.~Hu,
S.~Jin,
J.~Kile,
P.A.~McNamara III,
J.~Nielsen,
Y.B.~Pan,
\mbox{J.H.~von~Wimmersperg-Toeller}, 
W.~Wiedenmann,
J.~Wu,
Sau~Lan~Wu,
X.~Wu,
G.~Zobernig
\nopagebreak
\begin{center}
\parbox{15.5cm}{\sl\samepage
Department of Physics, University of Wisconsin, Madison, WI 53706,
USA$^{11}$}
\end{center}\end{sloppypar}
\vspace{2mm}
\begin{sloppypar}
\noindent
G.~Dissertori
\nopagebreak
\begin{center}
\parbox{15.5cm}{\sl\samepage
Institute for Particle Physics, ETH H\"onggerberg, 8093 Z\"urich,
Switzerland.}
\end{center}\end{sloppypar}
}
\mbox{ }\\
\vspace{-8.5cm}
\footnotetext[1]{Also at CERN, 1211 Geneva 23, Switzerland.}
\footnotetext[2]{Now at LAPP, 74019 Annecy-le-Vieux, France}
\footnotetext[3]{Also at Dipartimento di Fisica di Catania and INFN Sezione di
 Catania, 95129 Catania, Italy.}
\footnotetext[4]{Deceased.}
\footnotetext[5]{Also Istituto di Cosmo-Geofisica del C.N.R., Torino,
Italy.}
\footnotetext[6]{Now at Institut f\"ur Experimentelle Kernphysik, Universit\"at Karlsruhe, 76128 Karlsruhe, Germany.}
\footnotetext[7]{Supported by CICYT, Spain.}
\footnotetext[8]{Supported by the National Science Foundation of China.}
\footnotetext[9]{Supported by the Danish Natural Science Research Council.}
\footnotetext[10]{Supported by the UK Particle Physics and Astronomy Research
Council.}
\footnotetext[11]{Supported by the US Department of Energy, grant
DE-FG0295-ER40896.}
\footnotetext[12]{Now at Departement de Physique Corpusculaire, Universit\'e de
Gen\`eve, 1211 Gen\`eve 4, Switzerland.}
\footnotetext[13]{Supported by the Commission of the European Communities,
contract ERBFMBICT982874.}
\footnotetext[14]{Also at Rutherford Appleton Laboratory, Chilton, Didcot, UK.}
\footnotetext[15]{Permanent address: Universitat de Barcelona, 08208 Barcelona,
Spain.}
\footnotetext[16]{Supported by the Bundesministerium f\"ur Bildung,
Wissenschaft, Forschung und Technologie, Germany.}
\footnotetext[17]{Supported by the Direction des Sciences de la
Mati\`ere, C.E.A.}
\footnotetext[18]{Supported by the Austrian Ministry for Science and Transport.}
\footnotetext[19]{Now at SAP AG, 69185 Walldorf, Germany}
\footnotetext[20]{Now at Groupe d' Astroparticules de Montpellier, Universit\'e de Montpellier II, 34095 Montpellier, France.}
\footnotetext[21]{Now at D\'epartement de Physique, Facult\'e des Sciences de Tunis, 1060 Le Belv\'ed\`ere, Tunisia.}
\footnotetext[22]{Supported by the US Department of Energy,
grant DE-FG03-92ER40689.}
\footnotetext[23]{Now at Skyguide, Swissair Navigation Services, Geneva, Switzerland.}
\footnotetext[24]{Also at Dipartimento di Fisica e Tecnologie Relative, Universit\`a di Palermo, Palermo, Italy.}
\footnotetext[25]{Now at CERN, 1211 Geneva 23, Switzerland.}
\footnotetext[26]{Now at Honeywell, Phoenix AZ, U.S.A.}
\footnotetext[27]{Now at INFN Sezione di Roma II, Dipartimento di Fisica, Universit\`a di Roma Tor Vergata, 00133 Roma, Italy.}
\footnotetext[28]{Now at Centre de Physique des Particules de Marseille, Univ M\'editerran\'ee, F-13288 Marseille, France.}
\footnotetext[29]{Also at Department of Physics, Tsinghua University, Beijing, The People's Republic of China.}
\footnotetext[30]{Now at SLAC, Stanford, CA 94309, U.S.A.}
\footnotetext[31]{Now at LBNL, Berkeley, CA 94720, U.S.A.}
\setlength{\parskip}{\saveparskip}
\setlength{\textheight}{\savetextheight}
\setlength{\topmargin}{\savetopmargin}
\setlength{\textwidth}{\savetextwidth}
\setlength{\oddsidemargin}{\saveoddsidemargin}
\setlength{\topsep}{\savetopsep}
\normalsize
\newpage
\pagestyle{plain}
\setcounter{page}{1}

\newpage
\pagestyle{plain}
\setcounter{page}{1}
%
%
\pagenumbering{arabic}
\normalsize
\setlength{\textheight}{23cm}
\setlength{\textwidth}{16cm}
\unitlength 1mm
\setlength{\topmargin}{-1cm}
\setlength{\oddsidemargin}{-0.5cm}
\setcounter{page}{1}

\section{Introduction}\label{sec:introduction}

 In November 2000, ten days after the closing down of the LEP
 collider, the ALEPH collaboration published the preliminary
 findings~\cite{aleph00} of their search for the Standard Model (SM)
 Higgs boson~\cite{higgs}.
 An excess of events was found in the data collected in the year 2000
 at centre-of-mass energies up to 209\,GeV, in agreement with the
 production of a 114--115\,$\Gcs$ SM Higgs boson. The probability
 that this excess is consistent with the background-only hypothesis
 was determined to be at the level of a few permil, corresponding to a
 $\sim 3\sigma$ effect. The other three LEP experiments have also
 reported their search results~\cite{delphi,opal,l3}.



In this letter, after a brief reminder in
Section~\ref{sec:methodology} of the overall analysis methodology, the
changes with respect to the preliminary analysis presented in
Ref.~\cite{aleph00} are described. These minor modifications mostly
affect the four-jet channel $\h\qq$, arising from the $\ee\r\h\Z$
Higgsstrahlung process with subsequent hadronic decays of the Higgs and
Z bosons, in which the three highest-purity events were selected. They
also affect, although to a lesser extent, the other three main
topologies, i.e., the missing energy channel $\h\nunu$, the leptonic
channel $\h\lplm$, where $\ell$ is either an electron or a muon, and
the final states with taus $\tptm\qq$, when either the Higgs or the Z boson
decays to $\tptm$.

The final updates to the analysis, described in
Section~\ref{sec:finaltouches} together with their effect on the
result, are fourfold:

\begin{itemize}
\item the data sample was reprocessed with the final detector calibration
and alignment constants for the year 2000;
\item the precise knowledge of the LEP centre-of-mass energy 
was propagated to the final results;
\item additional simulated event samples were produced for a statistically
more accurate prediction of the Standard Model backgrounds;
\item an algorithm was developed to reject beam-related backgrounds and was
applied in the four-jet channel.
\end{itemize}
The results of the final combination of the searches for the Standard
Model Higgs boson, with these updates included, are given in
Section~\ref{sec:sm-results}, followed by a complete discussion of the
systematic uncertainties in Section~\ref{sec:systematics}. Other
relevant details of the analysis can be found in Ref.~\cite{aleph00}.

The search for the $\bb\bb$ and $\bb\tptm$ final states, which may
arise from the associated production $\ee\r\h\A$ in two-Higgs doublet
models, was also updated in the framework of the Minimal
Supersymmetric extension of the Standard Model (MSSM) with the data
collected in the year 2000. The final combination of the hZ and hA
searches with the results obtained at lower
energies~\cite{aleph189,aleph202} is presented in
Section~\ref{sec:mssm-results}.

Finally, possible invisible decays of a Higgs boson produced via the
Higgsstrahlung process were investigated with the data collected in
the year 2000. The result of the combination with earlier
searches~\cite{aleph202,invhiggs} is reported in
Section~\ref{sec:inv-results}.


\section{Search methodology}\label{sec:methodology}


 In order to provide mutually cross-checked results, the Higgs boson
 search is carried out in two alternative ``streams'', the first
 relying mostly on neural networks (NN) for the event selections, and
 the second on sequential cuts. The final results are those obtained
 in the NN stream.
In the $\h\Z$ search, the two streams differ in the treatment of the
two most powerful search channels: the four-jet and the missing energy
final states. The treatment of the $\h\lplm$ and $\tptm\qq$ channels
is identical in the two streams. The defining characteristics of the
cut stream and of the NN stream are summarized in
Table~\ref{tab:streams}.

 The event selection criteria of the different search channels, used
 for the analysis of the 2000 data~\cite{aleph00}, are very similar to
 those used for the 1999 data~\cite{aleph202}. For the results
 presented in this letter the event selections are identical to those
 of Ref.~\cite{aleph00}, with only one improvement (described in
 detail in Section~\ref{sec:beam-bgds}) made to the four-jet
 selection.


 In each search channel the likelihood of a signal in the data is
 quantified by means of an extended likelihood ratio
 $Q$~\cite{glen-stats} 

\[
Q = \frac{L_{s+b}}{L_{b}} = \frac{e^{-(s+b)}}{e^{-b}}
\prod_{i=1}^{n_{\mathrm{obs}}} \frac{s f_{s}(\vec{X}_{i}) + b f_{b}(\vec{X}_{i})}
{b f_b(\vec{X}_{i})},
\]
which combines information about the numbers of events observed
($n_{\mathrm{obs}}$) and expected in both the background-only ($b$)
and the signal ($s+b$) hypotheses.~It also contains information,
through the signal and background probability density functions
(pdf's) $f_s$ and $f_b$, that provides additional discrimination
between the two hypotheses.  The pdf's are evaluated for each observed
candidate $i$, with measured discriminant variables $\vec{X}_i$. The
discriminant variable(s) used in each search channel of the two
analysis streams are listed in Table~\ref{tab:streams}.  The
likelihood ratio for the combined search is the product of the
likelihood ratios of the individual search channels.

\begin{table}[htb]
\begin{center}
\caption{\small
The main features of the two analysis streams. For each search channel
($\h\qq,\,\h\nnbar$, etc.) the type of event selection (``Cuts'' or
``NN'') is indicated. The observables $X$ denote the discriminant
variables used for the calculation of the likelihood ratio: $\mrec$
denotes generically the reconstructed mass, as defined for the given
channel~\cite{aleph00}, and NN$_{\mathrm{output}}$ refers to the
output of the NN used for the event selection. \label{tab:streams}}
\vspace{0.3cm}
\begin{tabular}{|l|c|c|c|c|} \hline
& \multicolumn{2}{|c|}{Cut stream}
& \multicolumn{2}{|c|}{NN stream} \\ \hline
Search channel & Event     & Discriminant & Event     & Discriminant \\ 
               & selection & variable(s)  & selection & variable(s)  \\
\hline\hline
$\h\qqbar$ & Cuts & $X=\mrec$
& NN &  $\vec{X}=(\mrec$,\,NN$_{\mathrm{output}})$ \\ \hline
$\h\nnbar$ & Cuts & $X=\mrec$
& NN &  $\vec{X}=(\mrec,\b_{\mathrm {tag}})$  \\ \hline
$\h\lplm$ & Cuts & $\vec{X}=(\mrec,\b\tau_{\mathrm {tag}})$
          & Cuts & $\vec{X}=(\mrec,\b\tau_{\mathrm {tag}})$ \\ \hline
$\tptm\qqbar$ & NN & $X=\mrec$
& NN &  $X=\mrec$ \\ \hline
\end{tabular}
\end{center}
\end{table} 

 The cut stream uses mostly the reconstructed mass
 $\mrec$~\cite{aleph00}\,as a single discriminant. The exception is
 the $\h\lplm$ channel: in this case, as the event selection has no
 b-tagging cuts, the inclusion of the second discriminant (to tag b and
 $\tau$ jets) is necessary.


\section{Analysis updates}\label{sec:finaltouches}

The updates made to the analysis of Ref.~\cite{aleph00}, mentioned in
Section~\ref{sec:introduction}, are described in detail in the
following subsections. The effect of each of these updates on the
significance of the observed excess~\cite{aleph00} is displayed in
Table~\ref{tab:effects}. The properties of the most significant
four-jet candidates, after all the analysis updates are taken into
account, are listed in Table~\ref{tab:4jetcand}.

\begin{table}[thb]
\begin{center}
\caption{\small 
The successive effect of the analysis changes on the maximum
significance of the observed excess, for the two alternative analysis
streams.
\label{tab:effects}}
\vspace{0.3cm}
\begin{tabular}{|lcc|}
\hline
Update                  & Cut stream   & NN stream \\
\hline
\hline
Significance~\cite{aleph00} & $\phantom{+}3.06\sigma$   
                            & $\phantom{+}2.96\sigma$ \\
\hline
Final processing        & $+0.21\sigma$ & $-0.14\sigma$\\
LEP $\roots$            & $-$           & $-$ \\
Additional simulated & & \\
~event samples         & $-0.36\sigma$ & $-0.14\sigma$\\
Beam-background         & $+0.13\sigma$ & $+0.14\sigma$\\
\hline                                          
Final significance      & $\phantom{+}3.04\sigma$ & $\phantom{+}2.82\sigma$\\
\hline
\end{tabular}
\end{center}
\end{table}
\vspace{-1.0cm}
\begin{table}[thb]
\begin{center}
\caption{\small
Details of the five four-jet candidates selected with an event weight
greater than 0.3 at $\mh=115\,\Gcs$ in either the NN or cut
streams. Jets~3 and~4 are the Higgs boson jets. The weight
$w=\ln\left(1+{s}{f_s}/{b}{f_b}\right)$ of the candidates in each stream is also
given. For candidate $e$, the jet pairing shown is only selected in
the cut stream.
\label{tab:4jetcand}}
\vspace{0.3cm}
\begin{tabular}{|c|c|c|c|c|c|c|c|c|}
\hline
Candidate & $\mrec$ & 
\multicolumn{4}{c|}{b tagging} & Four-jet & 
$w_{\mathrm{NN}}$ & $w_{\mathrm{cut}}$\\
\cline{3-6}
(Run/Event) & (\Gcs) & Jet 1 & Jet 2 & Jet 3 & Jet 4 & NN & & \\
\hline
\hline
$a$ (56698/7455)& 109.9 & 0.999 & 0.831 & 0.999 & 0.197 & 0.999 & 0.59 & 0.25 \\
$b$ (56065/3253)& 114.4 & 0.996 & 0.663 & 1.000 & 0.996 & 0.997 & 1.44 & 0.81 \\
$c$ (54698/4881)& 114.1 & 0.124 & 0.012 & 0.998 & 0.999 & 0.997 & 1.76 & 0.61 \\
$d$ (56366/0955)& 114.4 & 0.201 & 0.051 & 0.998 & 0.956 & 0.933 & 0.41 & 0.62 \\
$e$ (55982/6125)& 114.4 & 0.071 & 0.306 & 0.449 & 0.998 & 0.687 &  --  & 0.63 \\
\hline
\end{tabular}
\end{center}
\end{table}
\vspace{-1.0cm}

\subsection{Final processing}

 The data were reprocessed with the final detector calibration and
 alignment constants.  This reprocessing resulted in the recovery of
 1\,${\rm pb}^{-1}$ of data.  The total integrated luminosity for the
 year 2000 is ${\cal{L}}=217.2\,\invpb$.

 The reprocessing can change by small amounts the value of measured
 event properties such as the reconstructed Higgs boson mass or the
 b-tagging probabilities. Events close to some of the selection cuts
 may therefore move into or out of the selected sample.
 About 95\% of the data events selected previously were also selected
 after the final processing. More specifically, the most signal-like
 events, i.e., those with a large contribution to the log-likelihood
 ratio $-2\ln\,Q$, are still selected after the final processing.

 In the cut-based four-jet channel, a new event is selected with a
 reconstructed Higgs boson mass of $111.8\,\Gcs$. Prior to the
 reprocessing, this event narrowly failed one of the b-tagging
 cuts. The two Higgs-candidate jets have b-tagging values of 0.870 and
 0.965, whereas the Z-candidate jets have b-tag values of 0.096 and
 0.277. (The output of the neural network b-tagging algorithm ranges
 from 0. for light-quark jets, to 1. for b-quark jets.) The missing
 energy of the event is 70 GeV and the total missing momentum is below
 $10\,\Gc$. A probable explanation for the large missing energy and low
 missing momentum is that two energetic neutrinos were produced almost
 back to back by two b-quark semileptonic decays. Indeed, in one of
 the b-tagged jets, an identified muon has a momentum of 1.7\,GeV/$c$
 transverse to the jet axis and is therefore consistent with a
 semileptonic decay of a b hadron. Another low-momentum muon is
 observed opposite to this jet, which further substantiates this
 hypothesis.
 
 In the cut stream, where only the reconstructed mass information is
 used as a discriminant, the event is assigned a weight
 $\ln(1+{s}{f_s}/{b}{f_b})=0.27$ at $\mh=115\,\Gcs$. In the NN stream,
 this event was already selected prior to the final processing. The
 event has a NN output of 0.90, and is therefore assigned a relatively
 low weight compared to the most significant
 candidates~\cite{aleph00}.

\subsection{LEP centre-of-mass energy}

 In the most recently available determination~\cite{lepecal}, the
 centre-of-mass energies are, on average, smaller than those used in
 Ref.~\cite{aleph00} by $\sim 140$\,MeV. When this effect is taken
 into account, the reconstructed Higgs boson masses of the candidate
 events are reduced by the same amount, and the number of signal
 events expected to be produced decreases from $10.1$ to $9.5$ for
 $\mh=115\,\Gcs$. The impact on the observed significance is
 negligible.

 \subsection{Additional simulated event samples}

 In order to further reduce the statistical uncertainty in the event
 selection efficiencies and in the pdf's, significantly larger event
 samples were generated. In particular, additional simulated
 background samples for the $\ee\r\bb(\gamma), \cc(\gamma), \WW$ and
 $\ZZ$ processes were generated at centre-of-mass energies $\sqrt{s}=$
 206.0, 206.7 and 207.0 GeV. The existing signal samples were also
 supplemented with samples of \ee\r\h\qq\ and \h\nunu\ events at
 $\sqrt{s}=206.7$ GeV. While most of these additional samples were
 used in the NN stream for the preliminary results~\cite{aleph00},
 they have only been included in the cut stream for this letter.

\subsection{Control of beam-induced backgrounds}\label{sec:beam-bgds}


 In one of the most significant four-jet events, called ``candidate
 {\it b}'' in Table~\ref{tab:4jetcand}, a 22\,GeV energy deposit is
 observed at small polar angle, in the plane of the collider.

 This deposit does not fit the hypothesis that it is part of the
 event. The total measured energy is considerably larger than
 $\sqrt{s}$ and the total measured momentum is aligned with that
 of the deposit.
 A reasonable kinematic fit quality is obtained only if this deposit
 is assumed to be extraneous to the event, i.e., produced by a
 beam-induced background particle.

\begin{figure}[htb]
\begin{picture}(170,80)
\put(0,0){\epsfig{file=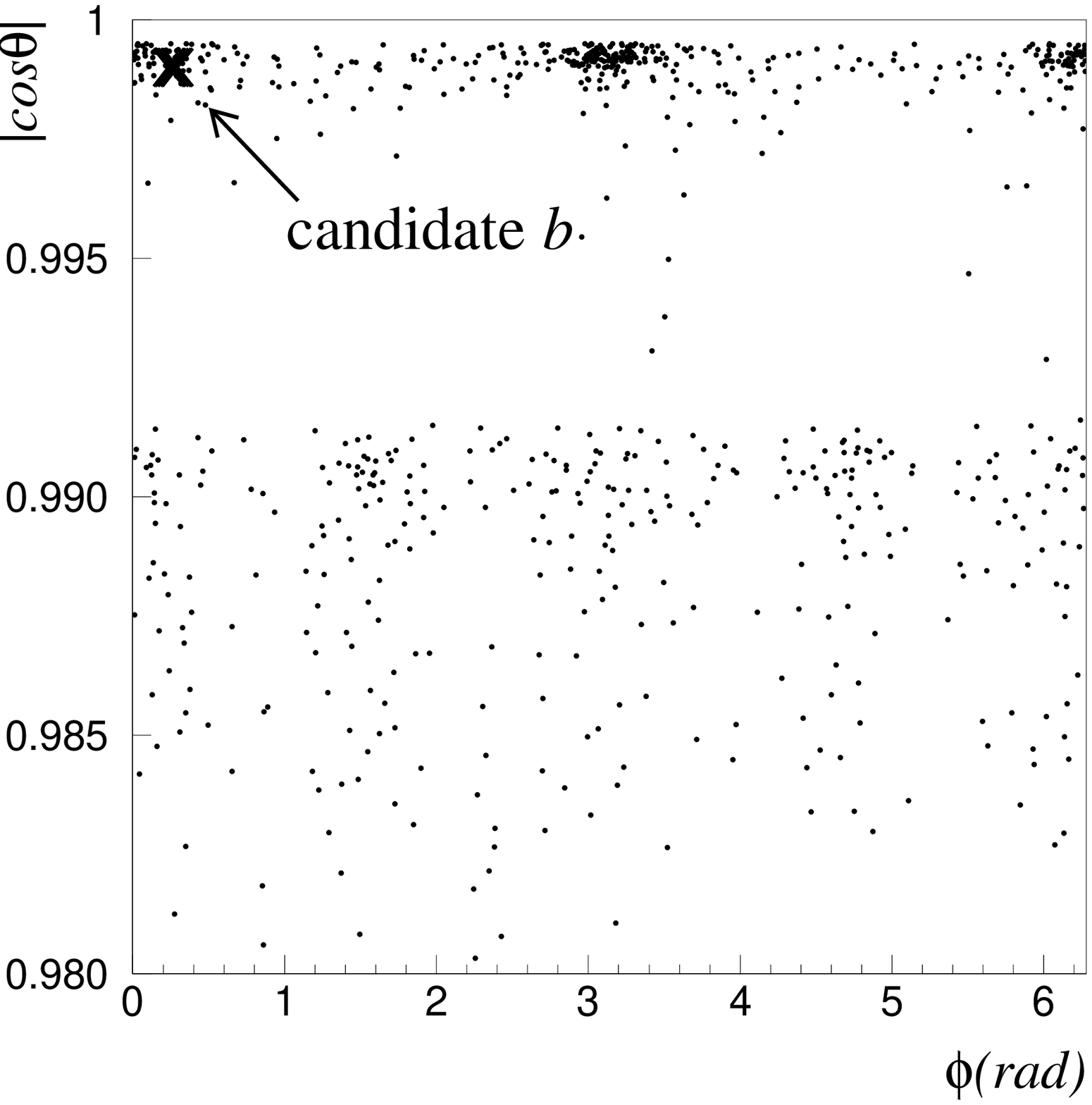,width=7.6cm}} 
\put(08,50){\Large \bf (a)}
\put(07,76){\Large ALEPH}
\put(85,0){\epsfig{file=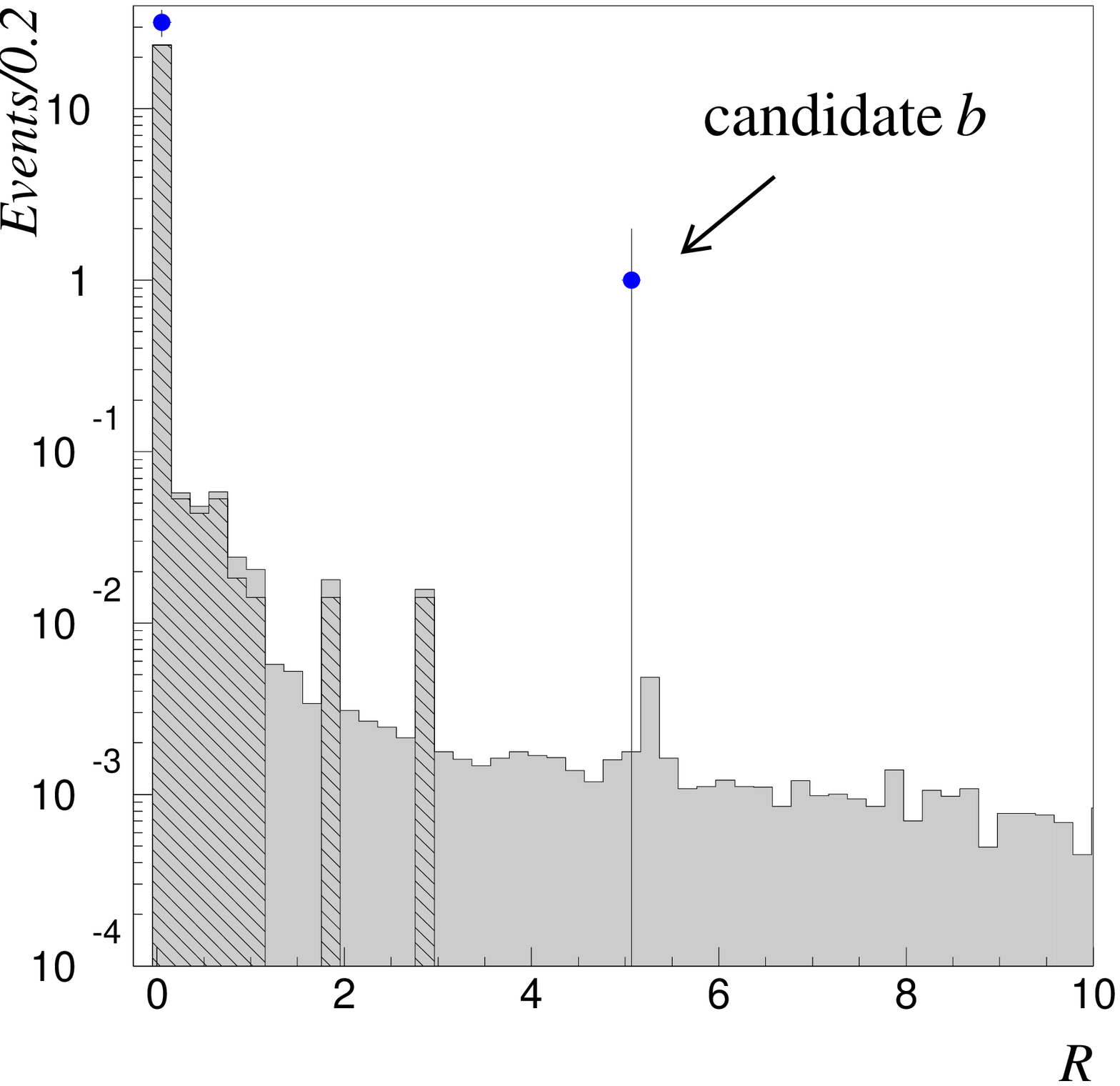,width=7.6cm}}
\put(96,50){\Large \bf (b)}
\put(92,76){\Large ALEPH}
\end{picture}
\caption{\small 
(a) The angular distribution of the most energetic cluster observed
within 12$^\circ$ of the beam axis, in a sample of events from
randomly-triggered beam crossings. Only clusters with $E>$\,3~GeV are
shown. The candidate event {\it b} is indicated by the cross in the
upper-left corner. The plane of the collider is defined by $\phi=0$
and $\pi$. (b) The distribution of $R$ for the events selected by the
cut-based four-jet search, for the expected SM background with (shaded
histogram) and without (hatched histogram) contamination from
beam-related background, and for the data (dots with error bars).
\label{fig:beam-bgd}}
\end{figure}

 It is indeed possible to observe large energy clusters from this
 background source. For example, in 0.89\% (0.48\%) of events
 triggered at random beam crossings, a deposit of energy in excess of
 3 (10) GeV is observed. The angular position of the most energetic
 cluster observed within 12$^\circ$ of the beam axis, in the
 randomly-triggered event sample, is shown in
 Fig.~\ref{fig:beam-bgd}a. The overwhelming majority of the
 beam-induced background particles are at very small polar angles and
 in the plane of the collider.

 As this type of background is not simulated, a procedure to
 identify and remove beam-background clusters had to be developed.
 The most energetic cluster with energy greater than 3 GeV, \mbox{$|\cos
 \theta\,|>0.998$} and which is isolated by at least \mbox{8$^\circ$}
 with respect to any other particle in the event, is fitted to each of
 the following three hypotheses.

\begin{itemize}
\item The identified cluster is part of the event.~In this case, 
the identified cluster is assigned to one of the four jets by the jet
clustering procedure.~The jets are subsequently fitted to the total
energy- and momentum-conservation constraints.

\item The identified cluster is, more specifically, 
assumed to be an ISR photon.~In this case, the rest of the event is
forced to form four jets.~These jets are fitted to the total energy-
and momentum-conservation constraints, modified to account for the
momentum imbalance caused by the hypothetical ISR photon.

\item The identified cluster is assumed to originate from a 
beam-induced background particle.~In this case too, the rest of the
event is forced to form four jets.~These jets are subsequently fitted
to the total energy- and momentum-conservation constraints.
\end{itemize}
 The $\chi^2$ values of these fits are henceforth designated
 $\chi_{\mathrm{norm}}^2, \chi_{\mathrm{ISR}}^2$ and
 $\chi_{\mathrm{beam}}^2$, respectively. The ratio

\[ R = \frac{\min(\chi_{\mathrm{norm}}^2,\chi_{\mathrm{ISR}}^2)}
{\chi_{\mathrm{beam}}^2}\] is expected to be larger for events
containing a beam-background particle. The distribution of $R$ for the
total expected SM background is shown in Fig.~\ref{fig:beam-bgd}b,
before and after it is ``contaminated'' with beam-background clusters
obtained from a sample of randomly-triggered events. Events in which
no energetic, isolated, small-angle cluster is found are assigned
$R=0$\,. Events with $R>2.$ are tagged as containing a beam-background
particle and the identified cluster of energy is removed from the
event prior to jet clustering and kinematic fitting. The remaining
events are treated according to the first hypothesis.

 The efficiency of the beam-background cleaning procedure,
 determined by running the algorithm on a contaminated background
 sample, is 28\% (50\%) for events with energy deposits in excess of
 3 GeV (10 GeV). The purity of the identification procedure is close
 to 100\%.

 At the final selection level 1.2\% of the simulated events are
 affected (i.e., newly selected, no longer selected, or with an
 $M_{\rm rec}$ value changed by at least 1\,GeV/$c^2$) by the
 contamination. This fraction is reduced to 0.4\% after the cleaning
 procedure is applied. The corresponding changes to the selection
 efficiencies are statistically insignificant and the changes in the
 pdf's imperceptible.

 When applied to the data, the cleaning algorithm identifies only one
 event (candidate {\it b}) as containing a beam-induced energy
 deposit. The deposit is therefore ignored in the analysis of this
 event, and the reconstructed Higgs mass (neural network output)
 changes from $112.8\,\Gcs$ (0.996) to $114.4\,\Gcs$ (0.997).


\section{Results of the SM Higgs search } \label{sec:sm-results}

In the 217.2\,${\rm pb}^{-1}$ of data collected during the year 2000, 
137 (99) events were selected in the NN (cut) stream, with 129.9 (88.2) 
expected from Standard Model backgrounds. The distribution of the events 
among the four search channels is shown in Table~\ref{tab:selected}. 
The mass distributions are very similar to those of Ref.~\cite{aleph00}.

\begin{table}[thb]
\begin{center}
\caption{\small 
The expected numbers of signal and background events and the numbers of
observed candidates in each search channel, for the two analysis
streams (``NN'' and ``Cut''). The signal expectation is
determined at $\mh=115\,\Gcs$.
\label{tab:selected}}
\vspace{0.3cm}
\begin{tabular}{|cccc|}
\hline
Search & Signal & Background & Events \\
 channel & expected & expected & observed \\
\hline
\hline
\h\qqbar\ (NN)  & 3.0   & 47.7  & 53 \\
\h\qqbar\ (cut) & 1.8  & 23.9  & 33 \\
\h\nnbar\ (NN)  & 1.0   & 37.7  & 39 \\
\h\nnbar\ (cut) & 0.9  & 19.8  & 21 \\
\h\lplm         & 0.4   & 30.8  & 30 \\
\tptm\qqbar & 0.3    & 13.7  & 15 \\ 
\hline
NN stream total         & 4.7   & 129.9 & 137 \\
Cut stream total        & 3.4   & 88.2  & 99 \\
\hline
\end{tabular}
\end{center}
\end{table}

\begin{figure}[htb]
\begin{picture}(170,80)
\put(0,0){\epsfig{file=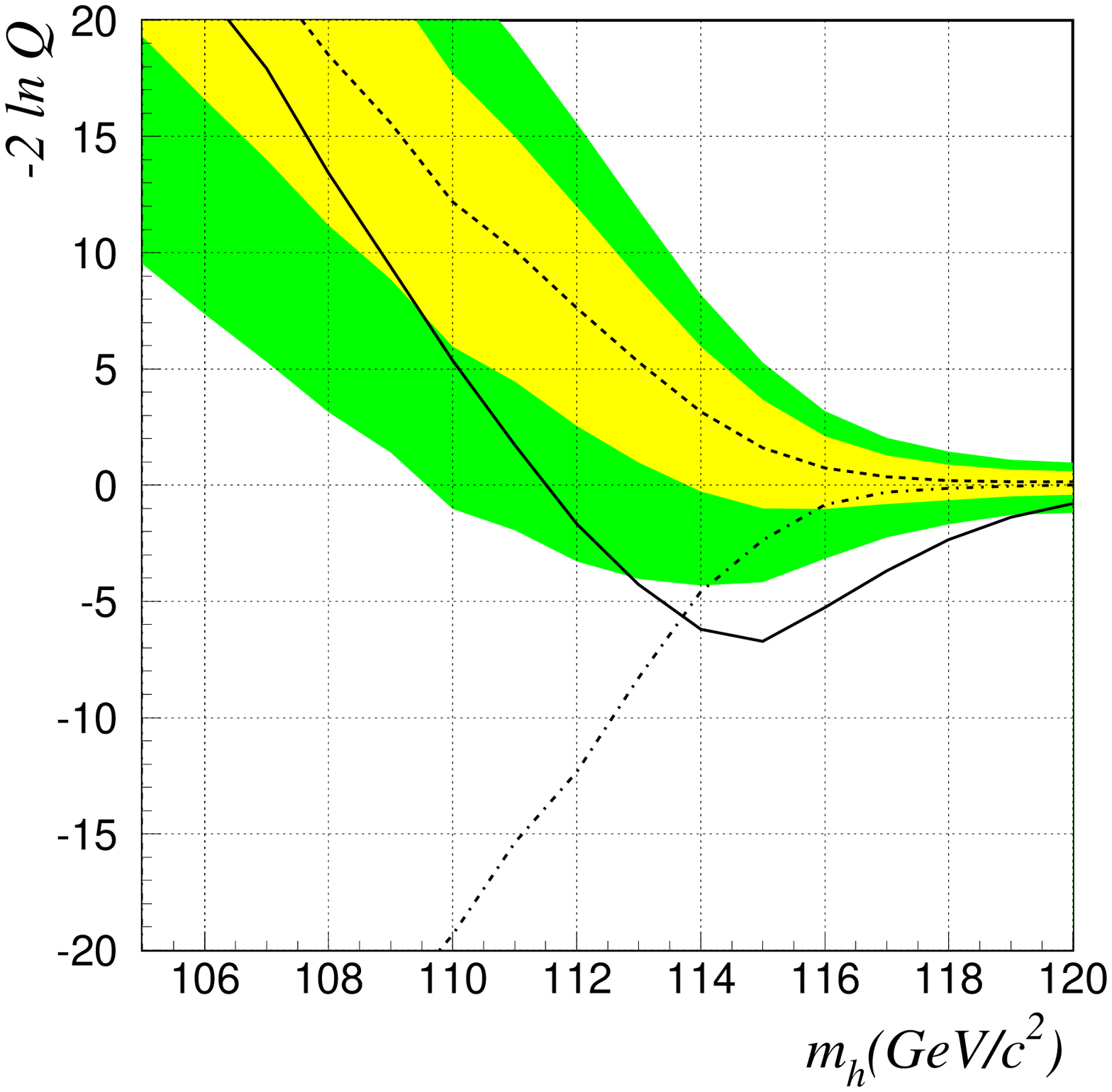,width=7.8cm}}
\put(60,63){\Large \bf (a)}
\put(10,80){\Large ALEPH}
\put(56,74){NN stream}
\put(85,0){\epsfig{file=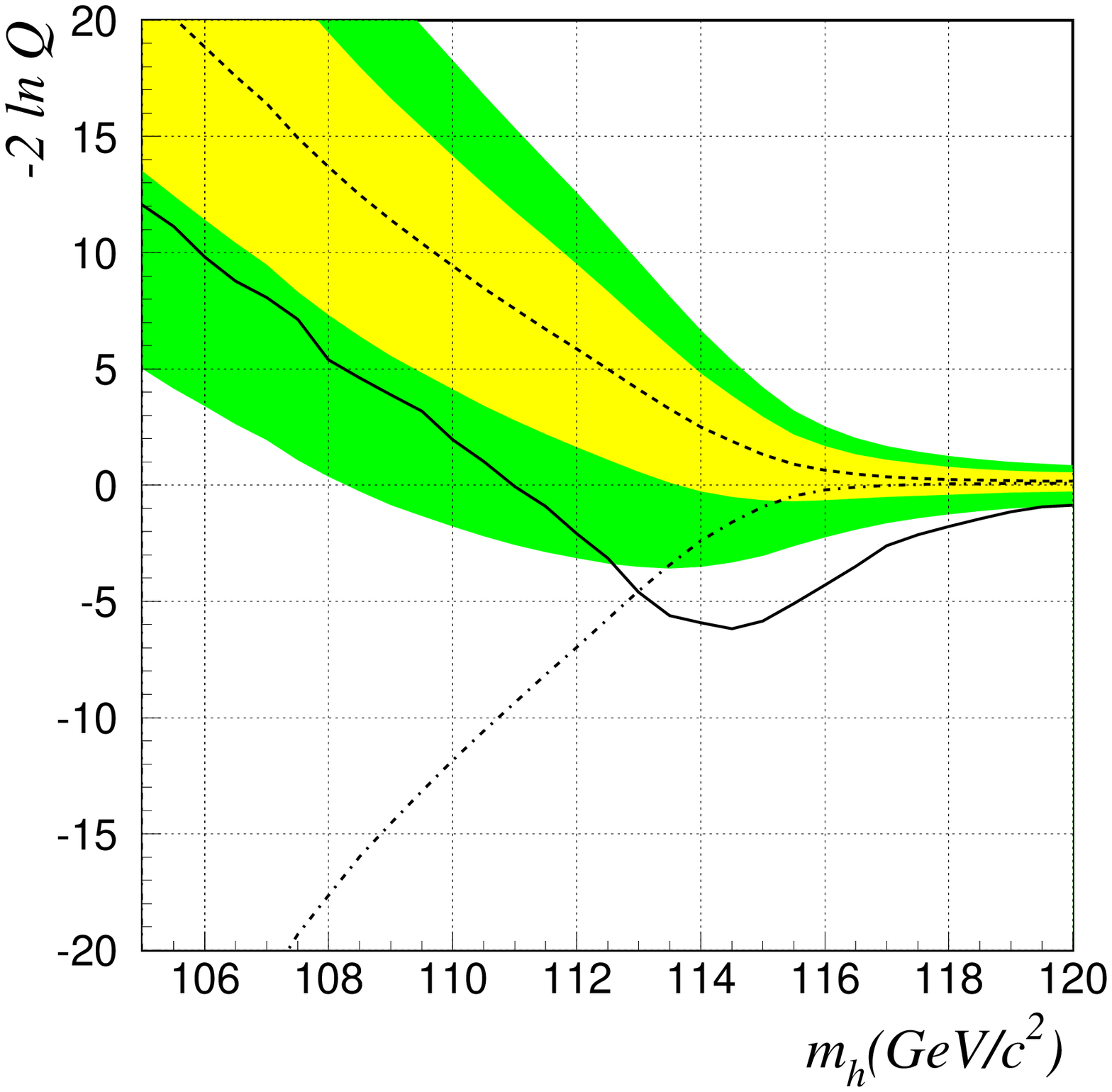,width=7.8cm}}
\put(144.5,63){\Large \bf (b)}
\put(95,80){\Large ALEPH}
\put(140,74){Cut stream}
\end{picture}
\caption{\small
The log-likelihood ratio, $-2\ln\,Q$, as a function of the test mass
$\mh$ for (a) the NN stream and (b) the cut stream, with all data
taken from 189 - 209 GeV. The solid line is the result obtained
from the data. The average result of background-only simulated
experiments is indicated by the dashed line; the light and dark shaded
bands around the background expectation contain 68\% and 95\% of the
simulated background-only experiments, respectively. The dash-dotted
curves indicate the expected position of the median log-likelihood
when the latter is calculated at a mass $\mh$ and includes a signal at
that same mass.
\label{fig:sm-lr}}
\end{figure}

The log-likelihood ratio, shown in Fig.~\ref{fig:sm-lr}a as a function
of the test mass $m_{\rm h}$, includes the data collected at smaller
centre-of-mass energies~\cite{aleph189,aleph202}. The large negative
values of the observed log-likelihood ratio indicate that the data
favour the signal hypothesis over the background-only hypothesis. The
most likely Higgs boson mass, corresponding to the minimum of
$-2\ln\,Q$, is around $\mh=115\,\Gcs$ for the NN stream. At this mass
the likelihood for the signal hypothesis, $L_{s+b}$, is 28.6 times
larger than the likelihood of the background-only hypothesis, $L_b$.
In the cut stream a similar result is observed
(Fig.~\ref{fig:sm-lr}b), with a preferred signal mass closer to
$\mh=114.5\,\Gcs$ and a factor 21.9 between the likelihoods of the two
hypotheses.  The probability (denoted $1-c_{\rm b}$) that this ratio
be even larger than observed in the background-only hypothesis is
shown in Fig.~4 as a function of $m_{\rm h}$. At the minimum of
$1-c_{\rm b}$, this probability is $2.4\times 10^{-3}$ ($1.1\times
10^{-3}$) in the NN (cut) stream, corresponding to an excess of 2.82
(3.04) standard deviations\footnote{The LEP Higgs working
group~\cite{lepcomb} has adopted a different convention, using a
double-sided Gaussian, to convert probability into standard
deviations. Under that convention the significance of the excess is
3.04 standard deviations in the NN stream and 3.25 standard deviations
in the cut stream.}. At the minimum of the likelihood ratio, the
significance of the excess is 2.70 (2.87) standard deviations. It is
consistent with the signal expectation at the 1.06 (1.29) standard
deviations level.
%

%
\begin{figure}[htb]
\begin{picture}(170,80)
\put(14,24){\Large \bf (a)}
\put(10,81){\Large ALEPH}
\put(58,75){NN stream}
\put(0,0){\epsfig{file=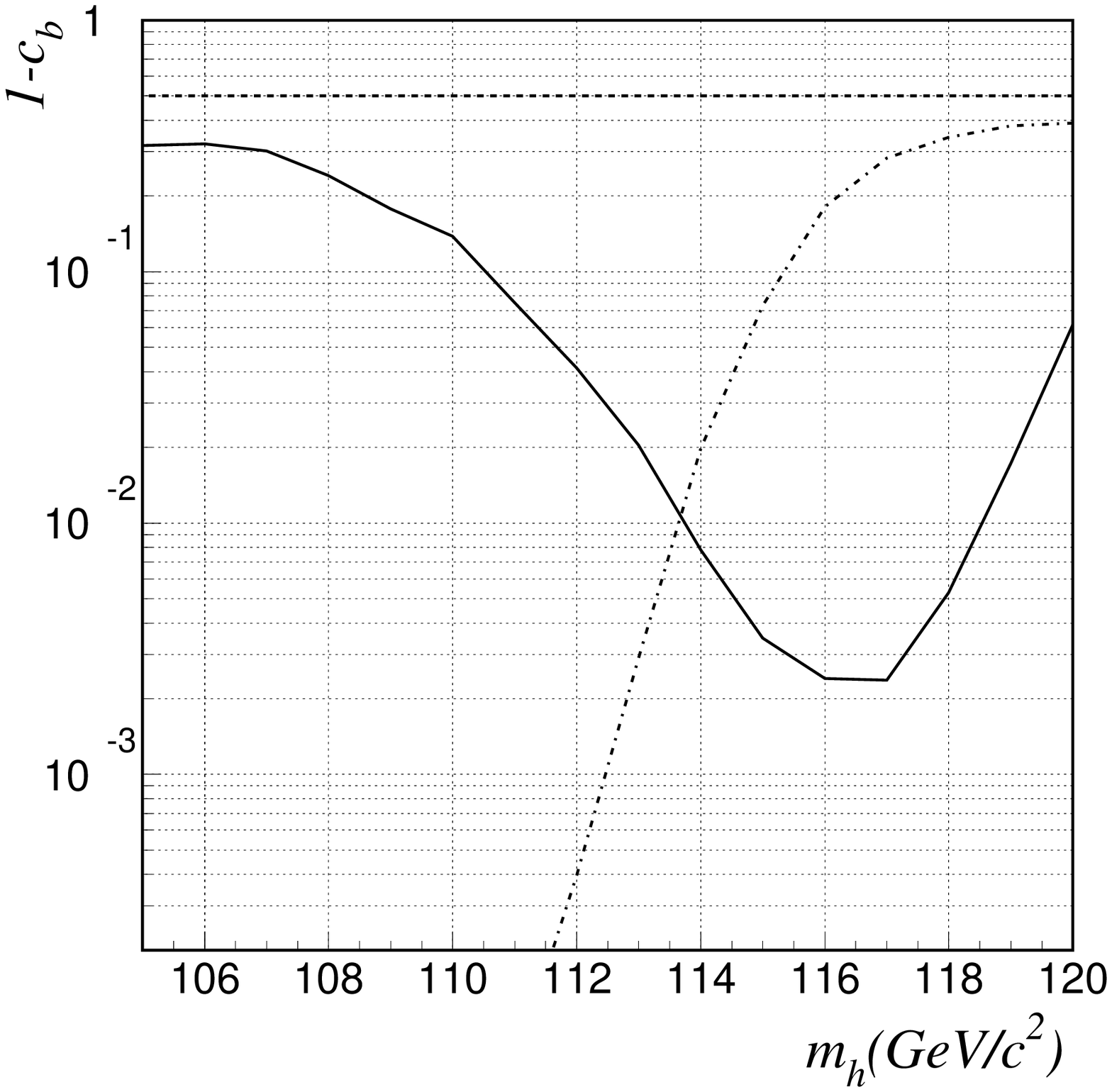,width=8.0cm}}
\put(99,24){\Large \bf (b)}
\put(95,81){\Large ALEPH}
\put(142,75){Cut stream}
\put(85,0){\epsfig{file=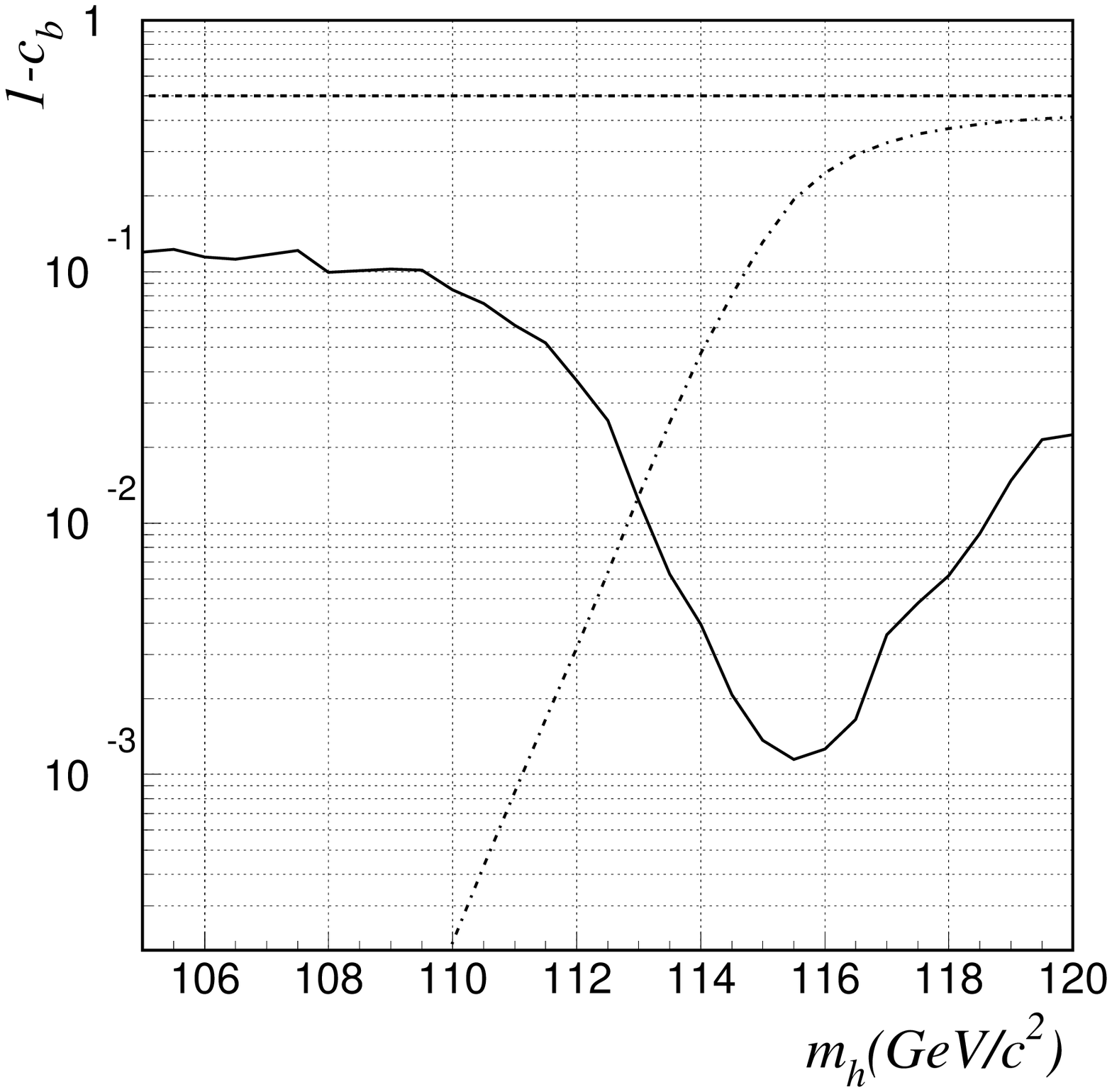,width=8.0cm}}
\end{picture}
\caption{\small
The observed (solid curve) and expected confidence
levels for the background-only (dashed curve) and the
signal (dash-dotted curve) hypotheses as a function of the
Higgs boson test mass for (a) the NN stream and (b) the cut
stream.
\label{fig:sm-cb}}
\end{figure}

Due to the observed excess, the 95\% C.L. lower limit of $111.5\,\Gcs$
set on $\mh$ in the NN stream search is well below the limit of
$114.2\,\Gcs$ expected in the absence of a signal. For comparison, a
lower limit of $110.4\,\Gcs$ is set in the cut stream, with
$113.6\,\Gcs$ expected.


\section{Systematic uncertainties}\label{sec:systematics}

 The results given in Section~\ref{sec:sm-results} include systematic
 uncertainties, incorporated according to the method of
 Ref.~\cite{cousins}. However, the significance of the excess might be
 affected by systematic uncertainties in a different manner. The
 systematic studies were therefore extended to estimate the impact of
 the dominant sources on the measured confidence levels
 (Fig.~\ref{fig:sm-cb}), especially around $\mh=116\,\Gcs$, where
 $1-c_b$ is smallest.
 The uncertainties in the $\h\nnbar$ channel were found to be negligible.
 The different systematic uncertainty sources and their impact on the
 observed significance are summarized in Table~\ref{tab:syst} and are
 discussed below. The uncertainties on the background dominate over
 those on the signal.

\begin{itemize}
\item[1.]{\bf{Statistics of simulated samples}}\\ To a large extent, 
the separation power between the background-only and the signal
hypotheses comes from the inclusion of the discriminant variable pdf's
in the likelihood ratio definition. It is especially so for a signal
close to threshold, for which the event rate is low. The statistical
uncertainty on the pdf's, which arises from the finite statistics of
the simulated samples, may therefore have an important impact on the
significance of the observed excess.

The small correlation between the two discriminant variables of the
NN-based four-jet search, ignored in the pdf's, was propagated to the
observed significance. The resulting correction is small, and its
uncertainty is limited by the finite statistics of the simulated
samples. The systematic uncertainty on the correction is estimated to
be half the size of the correction.

In the remaining search channels, the systematic uncertainty due to
the limited size of the simulated samples was determined by comparing
pdf's fitted to statistically independent samples of simulated
events. 
For instance, in the cut-based
four-jet channel, the estimated uncertainties in the pdf's in the high
reconstructed mass region are $\pm 1.5\%$ for the signal, and $\pm
10$--$20\%$ for the main background processes, $\ee\r\qq,\W\W$ and
$\Z\Z$.
The reconstructed mass pdf's were altered by $\pm 1 \sigma$ of these
estimated uncertainties. This alteration of the pdf's was applied to
the $3\,\Gcs$-wide region leading up to the kinematic threshold $\mrec =
\sqrt{s} - \mZ$, where the most significant candidates are
observed. Similarly, the reconstructed mass pdf's in the $\h\lplm$ and
$\tptm\qq$ channels were locally altered according to their estimated
uncertainty in the region $\mrec=116\,\pm\,2\,\Gcs$.
\pagebreak
\item[2.]{\bf{Tagging of b jets}} \\
 It has been determined~\cite{aleph189,aleph172} that the track impact
 parameter resolution is 5--10\% better in the simulation than in the
 calibration data taken at dedicated runs at the Z peak.  This is the
 main limitation of the simulation of the most relevant b-tagging
 distributions. The agreement between data and simulation is restored
 by smearing the track parameters in the simulation.

 The smearing effectively results in correcting the signal and
 background event selection efficiencies. The systematic uncertainty
 on the efficiencies was estimated to be half the size of the
 correction.
 The event selection efficiencies were therefore varied accordingly,
 under the assumption that the b-rich processes (e.g.,
 $\h\Z,\,\Z\Z,\,\Z\ee$) are fully correlated.
\item[3.]{\bf{Gluon splitting}} \\ The rate of gluon splitting 
to $\bb$ and $\cc$ quark pairs is underestimated in the simulation of
the $\ee\r\qq$ background. The measured splitting
rates~\cite{splitting} are enforced by reweighting the four-jet events
in the simulation that include a g$\r\bb$ or g$\r\cc$ branching. Twice
the uncertainty of these measurements was conservatively propagated to
the observed significance.
\item[4.]{\bf{Jet energy and angle resolutions}} \\ Small differences in the jet
energy resolution and jet energy scale are observed when comparing the
data and the simulation. The jet angular resolutions are also found to
be slightly better in the simulation.

The jets in the simulation were corrected~\cite{aleph189} to improve
the agreement between the simulation and the data, and a systematic
uncertainty amounting to half the size of the correction to the event
selection efficiencies was assumed.
\item[5.]{\bf{Simulation of other selection variables}} \\ The 
systematic effects potentially originating from event selection
variables other than those related to b tagging were evaluated with an
event reweighting method~\cite{aleph183}.  For each variable the event
weights were determined by making the simulated distribution agree
with that in the data at a preselection level, i.e., with ample
statistics. The effect of this reweighting on the selection
efficiencies was assumed to be due to a possible systematic
effect. Only small corrections, often statistically insignificant,
were found and their magnitude added in quadrature for all
variables.
\item[6.]{\bf{Strong coupling constant}}\\ A $\pm$5\% 
uncertainty on the strong coupling constant $\alpha_S$ was propagated
to the $\ee\r\qq$ background. 
\end{itemize}

\begin{table}[thb]
\begin{center}
\caption{\small
Variation in the significance of the observed excess in the two
analysis streams, at $\mh=116\,\Gcs$, due to the various systematic
error sources.
\label{tab:syst}}
\vspace{0.3cm}
\begin{tabular}{|lcc|}
\hline
Systematic &  &  \\
 source    & Cut stream & NN stream \\
\hline
\hline
Simulated statistics: 	& & \\
~-~$\tptm\qq$& $\pm 0.04\sigma$ 	& $\pm 0.02\sigma$ \\
~-~$\h\lplm$& $\pm 0.02\sigma$ 	& $\pm 0.02\sigma$ \\
~-~$\h\qq$ 	& $\pm 0.11\sigma$ 	& $\pm 0.07\sigma$ \\
%
Tagging of b jets & $\pm 0.06\sigma$ & $\pm 0.08\sigma$ \\
%
Gluon splitting & $\pm 0.04\sigma$ & $\pm 0.04\sigma$\\
%
Jet resolutions & $\pm 0.07\sigma$ & $\pm 0.05\sigma$ \\
%
Selection variables: &  &   \\
~-~$\h\lplm$ & $\pm 0.03\sigma$ & $\pm 0.03\sigma$ \\
~-~$\h\qq$ & $\pm 0.03\sigma$& $\pm 0.05\sigma$ \\
%
$\alpha_S$ & $\pm 0.04\sigma$ & $\pm 0.06\sigma$ \\
\hline
\end{tabular}
\end{center}
\end{table}

When all the uncertainties on the observed significance are added in
quadrature, the total systematic uncertainty is found to be $\pm
0.17\sigma$ for the cut stream and $\pm 0.15\sigma$ for the NN stream.

\section{Results in the context of the MSSM}\label{sec:mssm-results}

 In the MSSM, both the Higgstrahlung processes $\ee\r\h\Z$, with a
 cross section proportional to $\sba$, and the associated pair
 production $\ee\r\h\A$, with a cross section proportional to $\cba$,
 are searched for.  Here, $\tanb$ is the ratio of the vacuum
 expectation values of the two Higgs doublets and $\alpha$ is the
 Higgs mixing angle in the CP-even sector. As in the case of the SM
 Higgs boson search, the search for the MSSM Higgs bosons was also
 performed with the two alternative analysis streams~\cite{aleph202}. 


In the search for $\h\A$ pair production, ten events were selected in
the 2000 data in the $\bb\bb$ channel, with 5.5 events expected from
SM background processes. This slight excess of events is fully
correlated with that observed in the four-jet channel of the Standard
Model Higgs boson search. In the $\bb\tptm$ channel, three events
were selected with 3.0 events expected.

\begin{figure}[p]
\begin{center}
\begin{picture}(170,80)
\put(40,0){
\epsfig{file=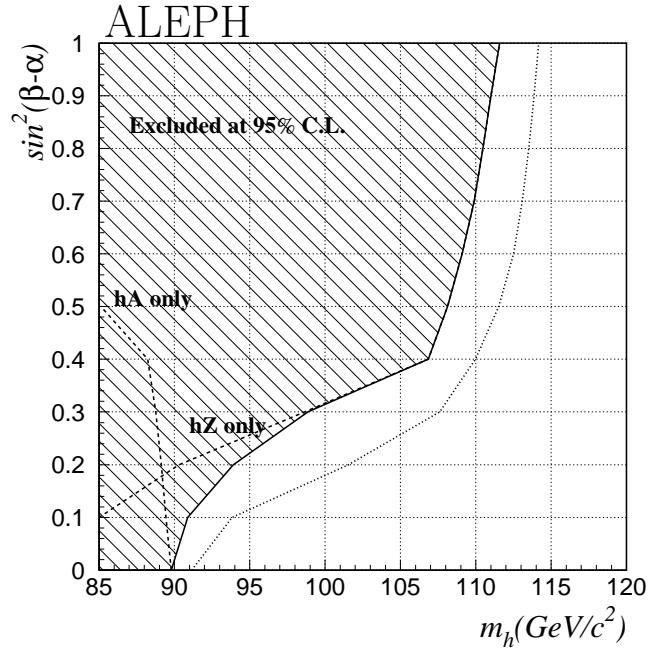,width=8.0cm}}
\put(52,80){\Large ALEPH}
\end{picture}
\vspace{-0.2cm}
\caption{\small 
The 95\% C.L. exclusion contours for the $\h\Z$ and $\h\A$
searches as a function of $\sba$ (dashed lines). The combined
exclusion is shown by the hatched area and the dotted line indicates
the expected exclusion.
\label{fig:mssm-sinab}}
\end{center}
\end{figure}

\begin{figure}[p]
\begin{center}
\begin{picture}(170,80)
\put(0,0){
\epsfig{file=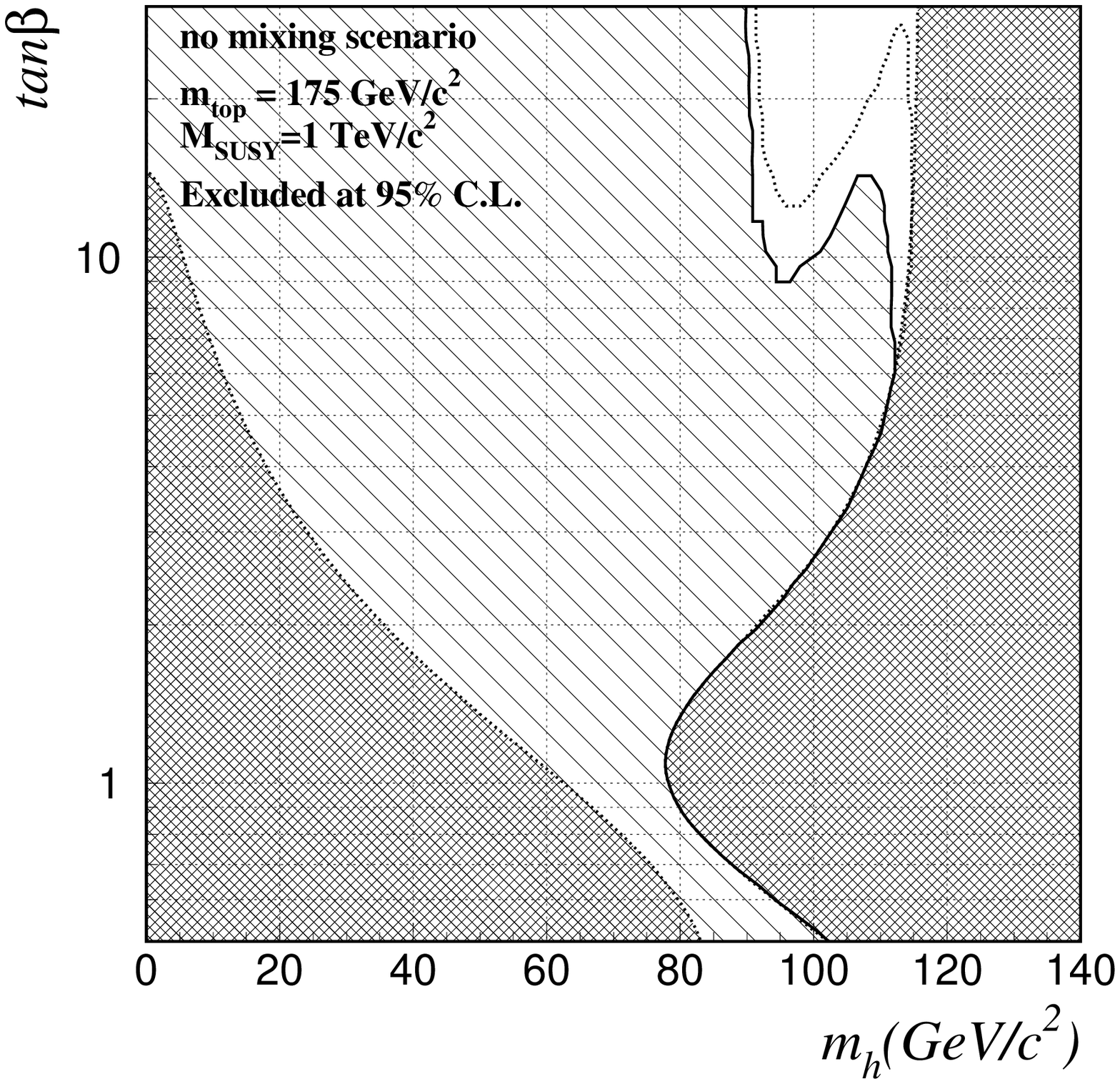,width=8.0cm}\hspace{0.8cm}}
\put(57,73){\Large \bf (a)}
\put(12,80){\Large ALEPH}
\put(85,0){
\epsfig{file=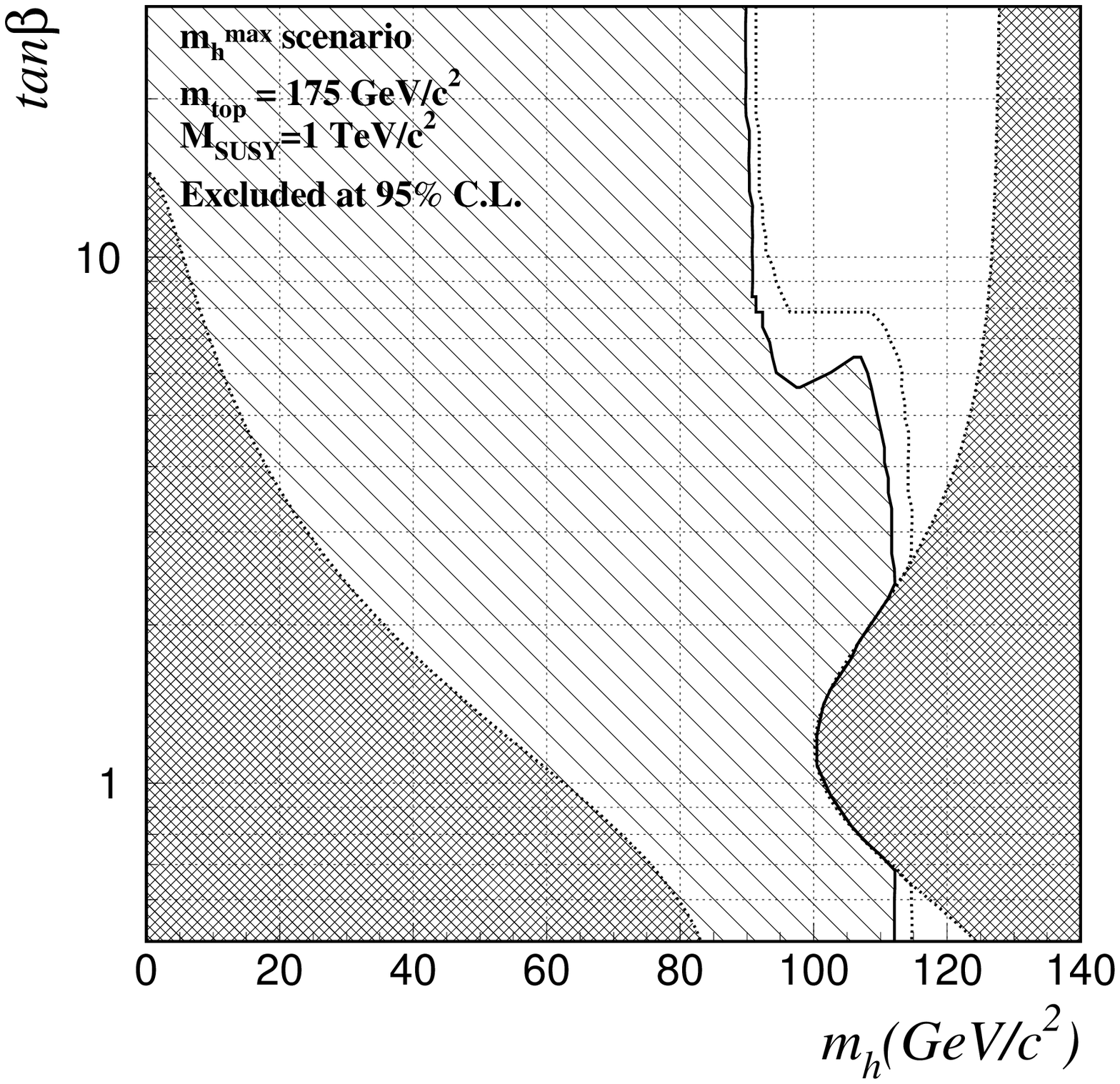,width=8.0cm}}
\put(147,73){\Large \bf (b)}
\put(97,80){\Large ALEPH}
\end{picture}
\vspace{-0.2cm}
\caption{\small
The experimentally excluded regions, at 95\% C.L., in the
$(\mh,\tanb)$ parameter space, for (a) the no-mixing and
(b) the \mhmax\ benchmark scenarios. The lightly-hatched
area is excluded experimentally. The dotted line indicates the
expected exclusion limit.The dark-hatched areas indicate
theoretically forbidden parts of the parameter space.
\label{fig:mssm-benchmarks}}
\end{center}
\end{figure}

The regions excluded at the 95\% C.L. by the hZ and the hA searches
independently, as well as by their combination, are shown in
Fig.~\ref{fig:mssm-sinab} as a function of $\sba$ with SM branching
fractions assumed for the lighter CP-even Higgs boson h. The combined
search allows an absolute lower limit on $\mh$ of 89.8\,GeV/$c^2$ to
be set at 95\% C.L.  These results are also interpreted in the context
of two MSSM benchmark scenarios, called ``no-mixing'' scenario and
``$m_{\rm h}^{\rm max}$'' scenario,
respectively~\cite{benchmarks}. The latter is expected to lead to
rather conservative $m_{\rm h}$ and $\tan\beta$ exclusions, while the
former is more favourable to LEP searches. The 95\% C.L. excluded
domains in the ($\mh, \tanb$) plane are shown for these two benchmark
scenarios in Fig.~\ref{fig:mssm-benchmarks}, with $\mtop=175\,\Gcs$.
The overall limits on $\mh$, $\mA$ and $\tanb$ are summarized in
Table~\ref{tab:mssm-limits}.

\begin{table}[htbp]
\begin{center}
\caption{\small
The excluded values (95\% C.L.) of $\mh,\,\mA$ and $\tanb$ in the two
MSSM benchmark scenarios described in the text. The numbers in
parentheses are the expected limits. The results are shown for the two
alternative analysis streams.
\label{tab:mssm-limits}}
\vspace{0.3cm}
\begin{tabular}{|c|cc|cc|}
\hline
&\multicolumn{2}{|c|}{NN stream} &\multicolumn{2}{c|}{Cut stream} \\
& No mixing & \mhmax\ & No mixing & \mhmax\ \\
\hline
\hline
$\mh<\,(\Gcs)$  & 89.8 (91.3)   & 89.8 (91.3) & 89.8 (90.8) & 89.8 (90.8)\\
$\mA<\,(\Gcs)$  & 90.1 (91.6)   & 90.1 (91.6) & 90.1 (91.3) & 90.1 (91.1)\\
$\tanb$ & [0.5--6.2]             & [0.7--2.3]   & [0.5--5.0] & [0.7--2.2] \\
\hline
\end{tabular}
\end{center}
\end{table}

 The theoretical upper limit on $\mh$ for a given $\tanb$
 (Figs.~\ref{fig:mssm-benchmarks}a, 5b) depends on $\mtop$. For
 $\mtop=180\,\Gcs$, the excluded $\tanb$ range is significantly
 reduced to [0.8, 1.8] in the $\mhmax$ scenario and to [0.5, 4.4] in
 the no-mixing scenario. The limits on $\mh$ and $\mA$ are not
 affected.


\section{Invisible Higgs boson search results}\label{sec:inv-results}

In models which allow the Higgs boson to decay invisibly, the
Higgstrahlung process gives rise to observable final states with
acoplanar lepton pairs ($\Z\r\lplm$) and with acoplanar jets
($\Z\r\qqbar$). An update of the searches~\cite{aleph202,invhiggs} for
these two topologies is presented in this section, with the data
collected in 2000.
                                                                                
The search for two acoplanar leptons was left unchanged; seven
events were selected, in agreement with 6.7 events expected
from background processes.
                                                                                
In the hadronic final state, the preselection was tightened to improve
the rejection of We$\nu$ events. The energy of the less energetic
hemisphere, formerly required to be nonzero~\cite{invhiggs}, is now
required to exceed 5$\%$ of the centre-of-mass energy. The data taken
in 1999~\cite{aleph202} were studied with a set of three NN-based
analyses, with the selection cut sliding as a function of the Higgs
boson mass hypothesis. Each neural network was optimized for a given
centre-of-mass energy (196, 200 or 202\,GeV). If the distance to
threshold $\sqrt{s} - \mh -\mZ$ is used as the sliding parameter
rather than the Higgs boson mass hypothesis, $\mh$, the networks need
neither be re-trained nor re-optimized. The same analysis can hence be
applied to the numerous centre-of-mass energies scanned in the year
2000 with nearly optimal neural network trainings and selection
criteria at each mass hypothesis. Altogether, 42 events were selected
in the data, compatible with the 48.6 events expected from background
processes.
                                                                               
\begin{figure}[htbp]
\begin{center}                                                                  
\begin{picture}(170,80)
\put(40,0){
\epsfig{file=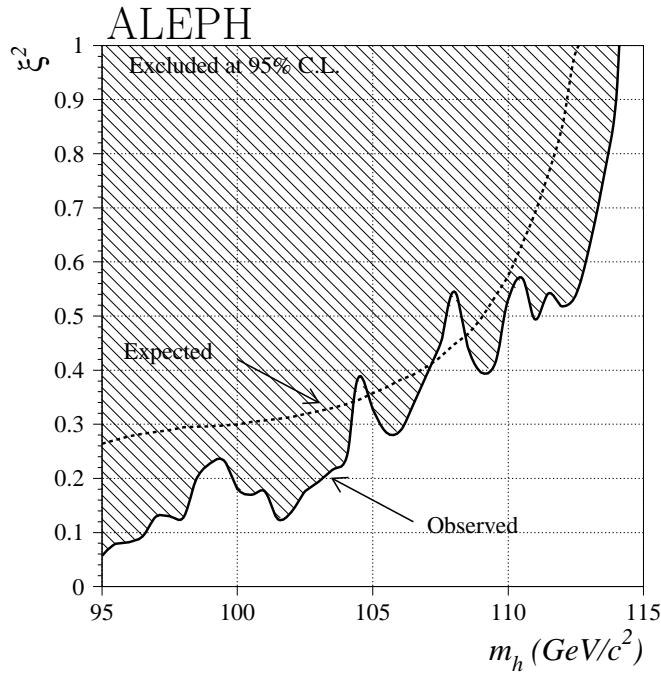,width=8.0cm}}
\put(50,80){\Large ALEPH}
\end{picture}
\vspace{-0.4cm}
\caption{\small
Result of the search for an invisibly-decaying Higgs boson. The observed 
(solid curve) and expected (dashed curve) exclusion regions in the 
(\mh, $\xi^2$) plane.} 
\label{plan}                                                                    
\end{center}                                                                    
\end{figure}

These results are interpreted as a lower limit on $\mh$ as a function
of $\xi^2$, the product of the invisible branching fraction of the
Higgs boson and a model-dependent factor which reduces the
Higgstrahlung cross section with respect to that in the Standard Model
(Fig.~\ref{plan}).  For $\xi^2 = 1$, the observed mass lower limit is
$114.1\,\Gcs$, for an expected 95\%~C.L. lower limit of $112.6\,\Gcs$.
\vspace{0.5cm}
%

\section{Conclusion}\label{sec:conclusion}

 The final results of the ALEPH search for the Standard Model Higgs
 boson have been presented and have been found to confirm the
 preliminary findings reported in the ALEPH
 publication~\cite{aleph00} that appeared shortly after the
 closing down of LEP.

 The analysis of all the data collected in the year 2000 up to
 centre-of-mass energies of 209\,GeV has been conducted with two
 parallel analyses, a neural-network-based stream and a cut-based
 stream. Both streams have revealed an excess with $\sim 3\sigma$
 significance, consistent with a Higgs signal around $115\,\Gcs$. The
 probability that such an excess is due to a fluctuation of the
 background is $2.4 \times 10^{-3}$ for the NN stream.
 Most of this effect arises in the four-jet search channel, as would
 be expected in the signal hypothesis. A detailed study of the most
 important systematic error sources has shown that the significance of
 the observed excess is robust. A 95\% C.L. lower limit on $\mh$ is
 set at $111.5\,\Gcs$.

 In the framework of the MSSM, the searches for the hZ and hA
 processes have allowed absolute lower limits of 89.8 and 90.1\,$\Gcs$
 to be set at 95\% C.L. on the h and A masses, and the range $0.7 <
 \tanb < 2.3$ to be excluded if $\mtop=175\,\Gcs$. The search for an
 invisibly decaying Higgs boson in hZ production has allowed a 95\%
 C.L. lower limit on $\mh$ to be set at 114.1\,$\Gcs$, for a cross
 section equal to that in the Standard Model and a 100\% branching
 fraction to invisible decays.



\section*{Acknowledgements}
We congratulate our colleagues from the accelerator divisions for the
very successful running of \LEP\ at high energies.  Without the
extraordinary achievement of operating LEP at energies much above the
design value, these observations would not have been possible.  We are
indebted to the engineers and technicians in all our institutions for
their contribution to the excellent performance of \ALEPH.  Those of
us from non-member countries thank \CERN\ for its hospitality.


\begin{thebibliography}{99}

\bibitem{aleph00} 
The ALEPH Collaboration, {\it{Observation of an excess in
the search for the Standard Model Higgs boson at ALEPH}},
\PLB{495}{2000}{1}.

\bibitem{higgs}
P.W. Higgs, Phys. Lett. {\bf 12} (1964) 132;
Phys. Rev. Lett. {\bf 13} (1964) 508; Phys. Rev. {\bf 145} (1966) 1156;\\
F. Englert and R. Brout, Phys. Rev. Lett. {\bf 13} (1964) 321;\\
G.S. Guralnik, C.R. Hagen, and T.W.B. Kibble, Phys. Rev. Lett.
{\bf 13} (1964) 585;\\
T.W.B. Kibble, Phys. Rev. {\bf 155} (1967) 1554.


\bibitem{delphi}
The DELPHI Collaboration, {\it Search for the Standard Model
Higgs boson at LEP in the year 2000}, \PLB{499}{2001}{23}. 
%
\bibitem{opal}
The OPAL Collaboration, {\it Search for the Standard Model
Higgs boson in $e^+e^-$ collisions at $\sqrt{s}\approx\,$192--209\,GeV},
\PLB{499}{2001}{38}.
%
\bibitem{l3}
The L3 Collaboration, {\it Standard Model Higgs boson with
the L3 experiment at LEP}, \PLB{517}{2001}{319}.
%
\bibitem{aleph189} 
The ALEPH Collaboration, {\it{Search for the neutral Higgs
bosons of the Standard Model and the MSSM in $e^+e^-$ collisions at
$\sqrt{s}$ = 189 GeV}},
\EPJ{17}{2000}{223}.
%
\bibitem{aleph202}
The ALEPH Collaboration, {\it{Searches for neutral Higgs
bosons in $e^+e^-$ collisions at centre-of-mass energies from 
192 to 202 GeV}}, \PLB{499}{2001}{53}.
%
\bibitem{invhiggs}
The ALEPH Collaboration,  
{\it Search for an invisibly decaying Higgs boson in $e^+e^-$
collisions at 189~GeV}, \PLB{466}{1999}{50}.
%
%
%
\bibitem{glen-stats}
see e.g., G. Cowan, {\it{Statistical Data Analysis}}, Oxford
University Press, 1998.
%
\bibitem{lepecal}
The LEP Energy Working Group, {\it{Evaluation of the LEP
centre-of-mass energy for data taken in 2000}}, LEP Energy Group
01-01.
%
\bibitem{lepcomb}
The LEP working group for Higgs boson searches, with the ALEPH,
DELPHI, L3 and OPAL Colls., {\it Search for the Standard Model Higgs
boson at LEP}, CERN-EP/2001-055 (July 2001).
%
%
\bibitem{cousins}
R.D. Cousins and V.L. Highland, {\it{Incorporating systematic
uncertainties into an upper limit}}, \NIM{320}{1992}{331}.
%
%
\bibitem{aleph172}
The ALEPH Collaboration, {\it Search for the neutral Higgs
bosons of the MSSM in $e^+e^-$ collisions at $\sqrt{s}$ from 130 to 172 GeV,}
\PLB{412}{1997}{173}.
%
\bibitem{splitting}
The LEP Heavy Flavour working
group, {\it{Input parameters for the LEP/SLD electroweak heavy
flavour results for Summer 1998 conferences}}, LEP-HF/98-01.
%
\bibitem{aleph183} 
The ALEPH Collaboration,  
{\it{Search for the Standard Model Higgs boson at the LEP2 collider
near $\sqrt{s}$ = 183 GeV}},
\PLB{447}{1999}{336.}
%
%
%
\bibitem{benchmarks}
M. Carena, S. Heinemeyer, C.E.M. Wagner and G. Weiglein, {\it
Suggestions for improved benchmark scenarios for Higgs boson searches
at LEP2}, CERN-TH/99-374 (1999).
%
\end{thebibliography}
\end{document}